\documentstyle[12pt]{article}

\advance\voffset by -1.5cm
\advance\hoffset by -1.5cm
\def\be{\begin{equation}}
\def\ee{\end{equation}}

\def\Zop{\bbbz}

\def\pmb#1{\setbox0=\hbox{#1}%
 \kern-.025em\copy0\kern-\wd0
 \kern.05em\copy0\kern-\wd0
 \kern-.025em\raise.0433em\box0 }

\def\H{{\cal H}}
\def\F{{\cal F}}
\def\G{{\cal G}}

\def\3{\ss}
\def\sq{\hbox{\rlap{$\sqcap$}$\sqcup$}}
\def\qed{\ifmmode\sq\else{\unskip\nobreak\hfil
\penalty50\hskip1em\null\nobreak\hfil\sq
\parfillskip=0pt\finalhyphendemerits=0\endgraf}\fi}
\def\half {\frac{1}{2}}

\def\thalf {\frac{3}{2}}

\def\bbbz {{\sf Z\!\!Z}}
\def\bbbr {{\rm I\!R}}

\begin{document}
\renewcommand{\theequation}{\thesection.\arabic{equation}}

\thispagestyle{empty}
\def\thefootnote{\fnsymbol{footnote}}
\begin{flushright}
  hep-th/9806024\\
  DAMTP 98-37
\end{flushright}
\vskip 2.5em
\begin{center}\LARGE
   The No-ghost Theorem for AdS${}_3$ and the Stringy Exclusion Principle
\end{center}\vskip 2em
\begin{center}\large
  Jonathan M. Evans%
  \footnote{Email: {\tt J.M.Evans@damtp.cam.ac.uk}}%
  \hskip0.5em Matthias R. Gaberdiel%
  \footnote{Email: {\tt M.R.Gaberdiel@damtp.cam.ac.uk}}%
  \hskip0.5em and 
  Malcolm J. Perry%
  \footnote{Email: {\tt M.J.Perry@damtp.cam.ac.uk}}%
\end{center}
\begin{center}\it
Department of Applied Mathematics and Theoretical
Physics, \\
University of Cambridge, Silver Street, \\
Cambridge CB3 9EW, U.K.
\end{center}
\vskip 1em
\begin{center}
 June 1998
\end{center}
\vskip 1em
\begin{abstract}
  A complete proof of the No-ghost Theorem for bosonic and
  fermionic string theories on ${\rm AdS}_3$, or the
  group manifold of $SU(1,1)$, is given. It is then shown that the
  restriction on the spin (in terms of the level) that is necessary to
  obtain a ghost-free spectrum corresponds to the stringy exclusion
  principle of Maldacena and Strominger. 
\end{abstract}

\setcounter{footnote}{0}
\def\thefootnote{\arabic{footnote}}

\section{Introduction}
\setcounter{equation}{0}

It has been conjectured recently that there exists a duality between
supergravity (and string theory) on $(D+1)$-dimensional anti-de Sitter
space, and a conformal field theory that lives on the $D$-dimensional
boundary of anti-de Sitter space \cite{M}. This proposal has been
further elaborated in \cite{GKP,Witten}, where a relation between the
correlation functions of the two theories has been proposed, and many
aspects of it have been analysed (see \cite{Witten1} and references
therein). 

In this paper we shall consider a specific example of the above class
of proposals, in which type IIB string theory on ${\rm AdS}_3 \times
S^3 \times M^4$ (where $M^4$ is either K3 or $T^4$) is conjectured to
be dual to a two-dimensional conformal field theory whose target
manifold is a symmetric product of a number of copies of $M^4$.  This
example is of special interest as both partners of the dual pair are
fairly well-understood theories, and and it should therefore be
possible to subject the proposal to non-trivial tests.  On the string
theory side, for instance, we have 
${\rm AdS}_3 \cong SL(2,\bbbr) \cong SU(1,1)$ and $S^3 \cong SU(2)$,  
and since these are both group manifolds we should be able to
determine the spectrum of string states exactly. The conjectured
duality together with the S-duality of type IIB relates this string
theory to a superconformal field theory with target space 
$Sym_{Q} M^4$, where $Q$ appears as the level of each WZW model, and
$Q$ is presumed to be large. It was pointed out in \cite{MS} that in
the dual conformal 
field theory, there are only finitely many chiral primary states, and
as a consequence this must be somehow reflected in the corresponding
string theory. This has lead to the proposal that there is a ``stringy
exclusion principle'' which removes certain states from the string
spectrum. It is the purpose of this paper to shed some light on this
proposal. In particular, we shall explain below how this restriction
has a natural interpretation in terms of the no-ghost-theorem for a
string theory on $SU(1,1)$. 
\medskip

The question of consistency of string theories on $SU(1,1)$ has a long
history \cite{BOFW}-\cite{Satoh}. There are effectively two different
approaches which are in a sense orthogonal to each other.
In the first approach, advocated some time ago 
by Hwang and collaborators 
\cite{Hwang1}-\cite{Hwang5} 
following the earlier work of \cite{Petr,Moh},
the Fock space of states that is analysed is the
space on which all generators of the Kac-Moody algebra of $su(1,1)$ are
well-defined, whereas this is not true in the second approach 
advocated by Bars \cite{Bars} and 
more recently Satoh \cite{Satoh} in which free-field-like Fock
spaces are introduced. 
These spaces are therefore 
at best different dense subspaces of the space of string states, 
and since very little is known about possible completions, it is not 
clear how these approaches are related, if at all. 

In this paper we shall follow the first approach, in which it is
necessary to restrict the set of $SU(1,1)$ representations to those
whose spin (in the case of the discrete series) is essentially bounded
by the level, as first proposed in \cite{Petr,Moh}. This restriction
guarantees that certain negative norm physical states are removed from
the spectrum, and it is this condition that we show to correspond to
the stringy exclusion principle.  Unfortunately, the various arguments
for the positivity of the physical states under this restriction that
have been given in the literature are not quite satisfactory: for
example, the ``proof'' in \cite{Hwang1,Hwang2} is clearly incomplete,
the restriction in \cite{Hwang2} is too strong, and there is a gap in
the proof in \cite{Hwang4}. We shall therefore give a complete
description of the proof. In the bosonic case, our argument follows
closely the approach of \cite{Hwang4} (which in turn follows the old
argument of Goddard \& Thorn \cite{GT}) together with the result of
\cite{DPL}. We then give what we believe to be the first correct
statement and proof of the corresponding result for the fermionic
case.
\smallskip

It should be stressed that these arguments only guarantee that the
string theory is free of ghosts at the free level. To get a consistent
(ghost free) interacting theory, it would be necessary to show that
crossing symmetric amplitudes can be defined whose fusion rules close
among the ghost-free representations. This is a rather difficult
problem as the fusion rules of the $SU(1,1)$ WZW model are not
well understood. On the other hand, one may regard the fact that this
theory with the appropriate truncation appears as the dual pair of a
very well understood conformal field theory as evidence that it is
indeed consistent.
\bigskip

The paper is organised as follows. In section 2, we describe our
conventions and give the proof of the no-ghost-theorem in the bosonic
case. Section 3 is devoted to a similar analysis of the fermionic
case. In section 4, we explain in detail that the bound that arises in
the no-ghost-theorem corresponds to the stringy exclusion principle. 
Section 5 contains our conclusions and open problems, and in the
appendix we give explicit examples of physical states for the
fermionic theory which demonstrate that the bound on the spin is
necessary to ensure positivity.

\section{The bosonic theory}
\setcounter{equation}{0}

\subsection{ $SU(1,1)$ WZW models and Strings}

We should first emphasize the intrinsic interest of string theory on
$SU(1,1) \cong SL(2,\bbbr)$ (quite independent of the spectacular
recent developments already mentioned). The standard procedure for
deciding whether a given string background is consistent is to check
for quantum conformal invariance of the world-sheet sigma-model, as
given by the vanishing of appropriate $\beta$-functions.  It is not
hard to see that these conditions are insensitive to some vital
properties, however: by these criteria a flat `spacetime' with 13
timelike and 13 spacelike directions would be a perfectly consistent
background for the bosonic string with $c=26$, and yet there will
clearly be physical states of negative norm in such a theory. If there
is a single time-direction, the no-ghost theorem \cite{B,GT} for the
bosonic string in flat Minkowski spacetime, Mink$_d$ ensures that
there are no negative-norm states for $d\leq 26$, and this can
immediately be  extended to backgrounds of the type 
Mink$_d \times {\cal M}$ with $2 \leq d \leq 26$ provided ${\cal M}$
corresponds to a unitary CFT of appropriate central charge. But if we
are considering a background whose geometry involves a time-like
direction in an essential way, then unitarity and the absence of
ghosts is something which must be scrutinized very carefully. 

To examine such issues it is natural to turn to the simplest string
models which one can hope to solve exactly, namely those for which the
backgrounds are group manifolds \cite{GepWit}.  If we require only a
single time-like direction then we are led to the non-compact group
$SU(1,1) \cong SL(2, \bbbr)$, or its covering space, as a laboratory
for testing these basic ideas about string theory \cite{BOFW}.  In
this section we shall consider string theory on $SU(1,1) \times {\cal
M}$ where ${\cal M}$ is some unspecified target space corresponding to
a unitary conformal field theory.  We now proceed to define the string
theory and its physical states in terms of an $SU(1,1)$ WZW model at
level $k$.

The Kac-Moody algebra corresponding to $su(1,1)$ is
defined by 
\be
[J^a_m,J^b_n] = i f^{ab}{}_{\!c} J^c_{m+n} + k m \eta^{ab}
\delta_{m,-n} \,, 
\ee
where $\eta^{ab}= diag(+1,+1,-1)$, and 
$$
f^{abc} \equiv f^{ab}{}_d \eta^{dc} = \varepsilon^{abc}\,.
$$
We can then define $J^{\pm}_n = J^1_n \pm i J^2_n$, and in terms of
these modes the commutation relations are
\be
\begin{array}{lcl}
{\displaystyle [J^+_m,J^-_n]} & = & {\displaystyle 
- 2 J^3_{m+n} + 2 k m \delta_{m,-n}}  \vspace*{0.1cm} \\
{\displaystyle [J^3_m,J^{\pm}_n]} & = & {\displaystyle
\pm J^\pm_{m+n}} \vspace*{0.1cm} \\
{\displaystyle [J^3_m,J^3_n]} & = & {\displaystyle
- k m \delta_{m,-n} \,.} 
\end{array}
\ee
The adjoint operator of $J^a_m$ is $J^a_{-m}$, and thus
\be
\left(J^{\pm}_m\right)^\ast = J^{\mp}_{-m} \qquad
\left( J^3_m \right)^\ast = J^3_{-m} \,.
\ee
The Sugawara expression for the Virasoro algebra is
\be
L_n = {1 \over 2(k-1)} \sum_l : \left[ 
{1\over 2} \left( J^+_{n+l} J^-_{-l} + J^-_{n+l} J^+_{-l}\right)
- J^3_{n+l} J^3_{-l} \right] : \,,
\ee
which satisfies the Virasoro algebra
\be
\label{vir}
[L_m,L_n]=(m-n) L_{m+n} + {c \over 12} m (m^2 -1) \delta_{m,-n} 
\ee
with 
\be
c= {3k \over k-1} \,.
\ee
Furthermore we have
\be
\label{virkm}
[L_n,J^{\pm}_m] = -m J^{\pm}_{n+m} \qquad
[L_n,J^3_m] = -m J^3_{n+m}\,.
\ee
In the following we shall always consider the case $k>1$, as then
$c>0$ and the group manifold has one time-like and two space-like
directions. 

The Kac-Moody algebra contains the subalgebra of zero modes $J^a_0$ 
for which we introduce the quadratic Casimir as
\be
Q = {1\over 2} (J^+_0 J^-_0 + J^-_0 J^+_0) - J^3_0 J^3_0 \,.
\ee
Representations of the $su(1,1)$ zero mode algebra are characterised by 
the value of $Q$ and $J^3_0$ on a cyclic state $|j,m\rangle$,
\be
Q |j,m\rangle = - j (j+1) |j,m\rangle \qquad
J^3_0 |j,m\rangle = m |j,m\rangle \,.
\ee
In the following we shall mainly be concerned with 
the unitary representations
$D_j^{-}$ of the $su(1,1)$ algebra for which a cyclic state can be
chosen to be of the form $|j,j\rangle$, where 
$j\in\{-1/2,-1,-3/2, \ldots\}$, and $J^+_0|j,j\rangle=0$. There
exists also another discrete series ($D_j^{+}$) whose cyclic state is
of the form $|j,-j\rangle$ with $j\in\{-1/2,-1,-3/2, \ldots\}$, and  
$J^-_0 |j,-j\rangle=0$. In addition, there exist the continuous
(unitary) series for which the states $|j,m\rangle$ have 
$j=-1/2+i\kappa$, and $m\in\Zop$ ($C_j^0$) or $m\in\Zop+1/2$
($C_j^{1/2}$); and also there exists the exceptional representations
with $-1/2\leq j <0$ and $m\in\Zop$. Finally, we should not forget the
trivial representation consisting of the single state 
with $j=m=0$.  

The only unitary representations of the $SU(1,1)$ group are those we
have listed above. If we consider the universal covering group of
$SU(1,1)$ however, then there are more general representations of the
type $D^{\pm}_j$ in which $j$ and $m$ need not be half-integral
(although the allowed values of $m$ within any irreducible
representation always differ by integers) and similarly there are
additional continuous representations where $m$ need not be
half-integral.  The group manifold of $SU(1,1)$ is topologically
$\bbbr^2 \times S^1$ (with the compact direction being timelike) and
this is responsible for the quantisation of $m$ in units of half
integers. By contrast, the simply-connected covering group is
topologically $\bbbr^3$.  As will become apparent in section~4, the
conjectured duality mentioned in the introduction would seem to
involve a string theory defined on $SU(1,1)$ itself, rather than on
its covering space, and so this is the case on which we shall
concentrate. This seems to be similar to what was found in the case of
${\rm AdS}_5$ in \cite{HO}. Our proof of the no-ghost theorem given below
applies equally well to either $SU(1,1)$ or its covering space,
however.
\medskip

For a bosonic string on $SU(1,1) \times {\cal M}$ the world-sheet
conformal field theory has a chiral algebra generated by two
commuting subalgebras: one is the Kac-Moody algebra
corresponding to $su(1,1)$, and the other subalgebra corresponds to a
unitary conformal field theory. The Virasoro generators of the whole
theory are then of the form $L_n = L^{su(1,1)}_n + L^0_n$, where 
$L^{su(1,1)}_n$ and $L^0_m$ commute. We shall consider the case where
the total conformal charge
$$
c = c^{su(1,1)} + c^0 = 26 \,,
$$
which is necessary for the BRST operator $Q$ to satisfy $Q^2=0$. Let
us denote by $\H$ the Fock space that is generated from the {\it
ground states} (that form a representation of the zero modes of the
whole theory) by the action of the negative modes. The conformal
weight of a ground state is $h^{su(1,1)} + h^0$, where $h^{su(1,1)}$
is the conformal weight of the $su(1,1)$ ground state representation,
and by the assumption on the unitarity of the commuting subtheory,
$h^0\geq 0$. The {\it physical states} in the Fock space are defined
to be those that satisfy the Virasoro primary condition 
\be
\label{vp}
L_n \psi = 0 \qquad n>0 \,,
\ee
and the mass shell condition
\be
\label{ms}
L_0 \psi = \psi \,.
\ee
Suppose then that the Casimir operator of the $su(1,1)$ ground state
representation takes the value $-j(j+1)$. If $\psi$ is a descendant at
grade\footnote{We shall use the terminology {\it grade} for what is
usually called {\it level} in string theory, and reserve the term 
{\it level} for the central term of the affine algebra.} $N$, the
second condition becomes 
\be
\label{ms1}
{-j (j+1) \over 2(k-1)} +N \leq 1 \,.
\ee
It follows immediately that for the continuous unitary representations
of $su(1,1)$, this condition can only be satisfied for $N=0$, as 
$-j(j+1) = 1/4+\kappa^2>0$; in this case all states satisfy
(\ref{vp}), and the norms are by construction unitary. We can
therefore concentrate on the discrete unitary
representations.\footnote{We shall always ignore the exceptional
representations, since they do not occur in the Peter-Weyl
decomposition of the $L^2$ space, and therefore should not contribute
in string theory.}

\subsection{The no-ghost theorem in the discrete case}

Let us now consider the case where the ground states transform
according to the discrete representation $D_j^-$ of $su(1,1)$. (The
case where the representation is $D_j^+$ can be treated similarly.)
It is easy to find states in the Verma module constructed from these
ground state representations which are Virasoro primary and satisfy
the mass-shell condition and yet which have negative norms for certain
values of $j$ and $k$ \cite{BOFW,Petr,Moh}. It would seem, therefore,
as though the no-ghost theorem fails in this case. Following the
proposal of \cite{Petr,Moh,Hwang1}, however, we shall show that if we
impose the additional restriction 
\be
\label{spinlevel}
0 < -j < k
\ee 
then all physical states in $\H$ indeed have positive norm. 
Notice that this restriction together with the mass-shell condition 
(\ref{ms}) implies severe restrictions on the allowed grades for
physical states. In particular since  
$j+1 > 1-k$ and $j<0$ we find that 
\be
\label{restriction}
N \leq 1 + {j(j+1) \over 2(k-1)} < 1 - {j \over 2} < 1 + {k \over 2} .
\ee
The last bound implies that for fixed level $k$, the physical states
only arise at a finite number of grades. (However, there are
infinitely many physical states at every allowed grade, since the
unitary representations of $SU(1,1)$ are infinite dimensional.)

Let us now turn to the proof of the no-ghost theorem, following the
general strategy of 
Hwang \cite{Hwang4} and \cite{GT}.   
We denote by $\F$ the subspace of $\H$ that is spanned by states
$\psi\in\H$ for which 
\be
J^3_n \psi = 0 \qquad L_n \psi = 0 \qquad \hbox{for $n>0$.}
\ee
We also denote by $\H^{(N)}$ the subspace of $\H$ that consists of 
states whose grade is less or equal to $N$. In a first step we want to
prove the following Lemma  
\bigskip

\noindent {\bf Lemma}. If $c=26$ and $ - k< j < 0$, the states of the
form 
\be
\label{basis}
|\{\lambda,\mu\},f\rangle:= L_{-1}^{\lambda_1} \cdots
L_{-m}^{\lambda_m} (J^3_{-1})^{\mu_1} \cdots (J^3_{-m})^{\mu_m}
|f\rangle \,, 
\ee 
where $f\in\F$ with $L_0 |f\rangle = h_f |f\rangle$ is at grade $L$
and $\sum_r r \lambda_r + \sum_s s \mu_s + L \leq N$, form a basis for
$\H^{(N)}$.
\medskip

\noindent {\bf Proof.} The proof proceeds in two steps. First, we 
prove that the states of the form (\ref{basis}) are linearly
independent. Let us define the Virasoro algebra corresponding to the
$U(1)$ theory generated by $J^3$ as 
\be
\label{L3}
L^3_n = - {1 \over 2k} \sum_m : J^3_{m} J^3_{n-m} : \,,
\ee
whose corresponding central charge is $c^3=1$. We can then
define 
\be
\label{coset}
L_n^c = L_n - L^3_n \,,
\ee
and by construction the $L_n^c$ commute with $J^3_m$, and therefore
with $L^3_m$, and define a Virasoro algebra with $c^c=25$. Using
(\ref{coset}), we can then rewrite the states of the form
(\ref{basis}) in terms of states where $L_r$ is replaced by
$L^c_r$. It is clear that this defines an isomorphism of vector
spaces, and it is therefore sufficient to prove that these modified
states are linearly independent. Since $L_n^c$ and $J^3_m$ commute,
the corresponding Kac-determinant is then a product of the
Kac-determinant corresponding to the $U(1)$ theory (which is always
non-degenerate), and the Kac-determinant of a Virasoro highest weight
representation with $c=25$ and highest weight 
\be
h^c = h_f + {m^2 \over 2k} \,,
\ee
where $h_f$ and $m$ are the $L_0$-eigenvalue and the $J^3_0$
eigenvalue of the state $|f\rangle$. If $f$ is at grade $M$, then
\begin{eqnarray}
h^c & = & - {j (j+1) \over 2(k-1)} + M + {m^2 \over 2k} + h^0
\nonumber \\
& = & - { j (k+j) \over 2k (k-1)} + {M (k+j) \over k} 
   - { j \over k} (j-m+M) + {1 \over 2k} (j-m)^2  + h^0\,,
\end{eqnarray}
and since $j<0$, $j+k>0$ and $j-m+M\geq 0$, and $h^0\geq 0$ it follows
that $h^c>0$. Since the only degenerate representations of the
Virasoro algebra at $c=25$ arise for $h\leq 0$ (see {\it e.g.}
\cite{cft}) it follows that the Kac-determinant is non-degenerate, and
the states of the form (\ref{basis}) are indeed linearly independent.

The final step, completing the proof, is to establish 
by induction on $N$ (as in \cite{GT}) that these states
form a basis of $\H^{(N)}$ for all $N\geq 0$. The induction start
$N=0$ is trivial. Suppose then that we have proven the statement for
$N-1$, and let us consider the states at grade $N$. Let us
denote by $\G^{(N)}$ the subspace of $\H^{(N)}$ that is generated by
the states of the form (\ref{basis}) with $L<N$. We have shown above
that $\G^{(N)}$ does not contain any null states, and this implies
that $\H^{(N)}$ is the direct sum of $\G^{(N)}$ and its orthogonal 
complement (in $\H^{(N)}$). By the induction hypothesis it follows
that every state in the orthogonal complement of $\G^{(N)}$ is
annihilated by $L_n$ and $J^3_n$ (with $n>0$), and therefore that the
orthogonal complement consists of states in $\F$. This completes the
proof of the Lemma.
\bigskip

Let us call a state {\it spurious} if it is a linear combination of
states of the form (\ref{basis}) for which $\lambda\ne 0$. Any given
physical state $\psi$ can then be written as a spurious state $\psi_s$
plus a linear combination of states of the form (\ref{basis}) with
$\lambda=0$, {\it i.e.}
\be
\psi = \psi_s + \chi \,.
\ee
For $c=26$, following the argument of Goddard and Thorn \cite{GT},
$L_1\psi_s$ and $\tilde{L}_2 \psi_s = (L_2 + 3/2 L_1^2)\psi_s$ are
again spurious states, and it follows that $\chi$ must also be a
physical state, {\it i.e.} that $L_n \chi=0$ for $n>0$. 
The next Lemma fills the gap in the argument given previously in
\cite{Hwang4}. 
\bigskip

\noindent {\bf Lemma}. 
Let $0>j>-k$. If $\chi$ is a physical state of the form (\ref{basis})
with $\lambda =0$, then $\chi\in\F$.
\medskip

\noindent {\bf Proof}. 
For fixed
$|f\rangle\in\F$, let us denote by  $\H_f$ the Fock space that is
generated by the action of $J^3$ from $|f\rangle$, and by $\H_f^{vir}$
the Fock space that is generated by the action of $L^3$ from
$|f\rangle$. Since $L^3$ can be expressed as a bilinear in terms of
$J^3$ (\ref{L3}), it is clear that  $\H_f^{vir}$ is a subspace of
$\H_f$. On the other hand  $\H_f^{vir}$ is a Virasoro Verma module for
$c=1$ whose ground state has conformal weight $- m^2 / 2k$ (where $m$
is the $J^3_0$ eigenvalue of $|f\rangle$), and it follows from the
Kac-determinant formula that $\H_f^{vir}$ does not contain any null
states unless $m=0$ \cite{cft}. Provided that $m\ne 0$, it is then
easy to see that $\H_f^{vir}$ and $\H_f$ contain the same number of
states at each grade, and this then implies that
$\H_f^{vir}=\H_f$. Since $\H_f^{vir}$ does not contain any null states
(with respect to the Virasoro algebra) it then follows that $\H_f$
does not contain any Virasoro primary states 
other than $|f\rangle$ itself. It therefore only remains to show that
all physical states have $m\ne 0$.  

The physical states at fixed grade $N$ form a representation under the
zero mode $su(1,1)$ algebra since $J^{\pm}_0\psi$ and
$J^3_0 \psi$ are physical states provided that $\psi$ is. If the
ground states form a representation $D_j^-$ of the $su(1,1)$ zero mode
algebra, then the possible representations at grade $N$ are of the
type $D_J^-$ with $J=j+N,j+N-1,\ldots,j-N$, and therefore $m\leq j+N$
for all physical states at grade $N$. 
To prove the lemma it therefore suffices to show
that the  mass shell condition (\ref{ms1}) together with $j+k> 0$
and $j<0$ implies that $j+N< 0$. 

Let us consider more closely those grades which are allowed by the 
mass-shell condition (\ref{ms}) and the spin-level restriction
(\ref{spinlevel}). If $0 > j > -1$ then the mass-shell condition
alone implies that $N=0$ is the only possibility. For 
$-1 \geq j \geq -2$ we claim that $N < 2$. To see this note that 
$N \geq 2$ implies $k \geq 2$ because of (\ref{restriction}). But then
$j(j+1) / 2(k-1) <1$ and so $N\geq2$ is still forbidden by
(\ref{restriction}). We have therefore shown that $j+N < 0$, as
required, if $0> j \geq -2$. But also if $j < -2$ (which allows 
$N \geq 2$) then we find that $j+N < 0$ directly from
(\ref{restriction}). This completes the proof. 
\bigskip

One may think that the Lemma should also hold under weaker
assumptions, but it is maybe worth mentioning that if there was no
restriction on $j$ and if $J^3$ was spacelike, the corresponding
statement would not hold: indeed there exists a state  
$[(J_{-1}^3)^2 - m J^3_{-2} ] | j,m\rangle$ with 
$m=\sqrt{{-k\over 2}}$ which is annihilated by all Virasoro positive
modes, but which is not annihilated by $J^3_2$.  
\bigskip

\noindent {\bf Theorem:} For $c=26$ and $0<-j<k$, every physical state
$\psi$ differs by a spurious physical state from a state in $\F$. 
Consequently, the norm of every physical state is non-negative.
\bigskip

\noindent{\bf Proof}.
This follows directly from the previous two lemmas and the 
fact that $\F$ is a subspace of the coset space 
corresponding to $su(1,1)/u(1)$ which has been shown to be 
unitary for $0>j>-k$ by Dixon {\it et.al.}
\cite{DPL}. 
\bigskip

We should mention that the above argument can also be used to give a
proof of the no-ghost theorem in the flat case. In this case, the
coset module is positive definite (without any restrictions on the
momenta) and only the calculations that demonstrate that $h^c$ is
positive and that the conformal weight of the ground state of
$\H_f$ is negative need to be modified. This can easily be done (see
also \cite{Hwang4}). 

Finally, since the norms of states based on $D^-_j$ 
are continuous functions of $j$ and $k$, the arguments above actually
show that the representations with $0 > j \geq -k$ do not
contain any negative norm physical states.\footnote{The same argument
cannot be applied to the limit $j=0$ however. At this value there are
new physical states at grade $N=1$ with $h^0=0$ (which are excluded by
the mass-shell condition for $j<0$), and the theorem does indeed fail:
the norms of the two physical states $J^+_{-1} |0\rangle$ and
$J^3_{-1} |0\rangle$ have opposite sign.} Furthermore, there are
certainly physical states with negative norm whenever $j<-k$
(see {\it e.g.} \cite{BOFW,Petr,Moh}), and so our result cannot be
improved.

\section{The supersymmetric theory}
\setcounter{equation}{0}

A fermionic string theory on $SU(1,1)$ is defined by a 
supersymmetric WZW model on this group manifold.
The supersymmetric Kac-Moody algebra corresponding to $su(1,1)$ is
generated by $J^{a}_n$ and $\psi^a_r$, where $a=\pm, 3$, $n\in\Zop$,
and $r$ is a half-integer in the NS sector (which we shall consider in
the following). The (anti-)commutation relations are 
\be
\begin{array}{lcl}
{\displaystyle [J^a_m, J^b_n]} & = & {\displaystyle
i f^{ab}{}_c J^c_{m+n} + k m \eta^{ab} \delta_{m,-n}} \vspace*{0.1cm}
\\ 
{\displaystyle [J^a_m, \psi^b_r]} & = & {\displaystyle
i f^{ab}{}_c \psi^c_{m+r}} \vspace*{0.1cm} \\
{\displaystyle \{\psi^a_r, \psi^b_s\}} & = & {\displaystyle
k \, \eta^{ab} \delta_{r,-s} \,,}
\end{array}
\ee
where $f^{ab}{}_c$ and $\eta^{ab}$ are the same structure constants and
metric, respectively as before for the bosonic case. We shall use the
metric (and its inverse) to raise (and lower) indices.

The universal algebra that is generated from $J^a$ and $\psi^a$ is
isomorphic to the direct (commuting) sum of a bosonic Kac-Moody
algebra and three free fermions. Indeed, if we define 
\be
\tilde{J}^a_m = J^a_m + {i \over 2k} f^{a}{}_{bc} \sum_r 
\psi^b_{m-r} \psi^c_r \,,
\ee
then 
\be
\begin{array}{lcl}
{\displaystyle [\tilde{J}^a_m, \tilde{J}^b_n]} & = & {\displaystyle
i f^{ab}{}_c \tilde{J}^c_{m+n} 
+ \tilde{k} m \eta^{ab} \delta_{m,-n}} \vspace*{0.1cm} \\
{\displaystyle [\tilde{J}^a_m, \psi^b_r]} & = & 0 \,,
\end{array}
\ee
where $\tilde{k}=k+1$. We can thus introduce a Virasoro algebra by the
Sugawara construction,
\be
L_m = {1 \over 2 (\tilde{k}-1)} \eta_{ab} 
\sum_l : \tilde{J}^a_{m-l} \tilde{J}^b_l :
+ {1 \over 2k} \eta_{ab} \sum_r r : \psi^a_{m-r} \psi^b_r : \,,
\ee
which satisfies  
\be
\label{stress}
\begin{array}{lcl}
{\displaystyle [L_m, J^a_n]} & = & {\displaystyle -n J^a_{m+n}} 
\vspace*{0.1cm} \\
{\displaystyle [L_m, \psi^a_r]} & = & {\displaystyle 
- \left( {m\over 2} + r \right) \psi^a_{m+r}} 
\end{array}
\ee
and the Virasoro algebra (\ref{vir}) with central charge 
\be
c = { 3 \tilde{k} \over \tilde{k} - 1} + {3\over 2} = 
    3 { 3 k + 2 \over 2 k} \,.
\ee
The theory has actually a super Virasoro symmetry, where the
additional generator is defined by 
\be
G_r = {1 \over k} \eta_{ab} \sum_s \tilde{J}^a_{r-s} \psi^b_s
 - {i \over 6 k^2} f_{abc} \sum_{s,t} \psi^a_{r-s-t} \psi^b_s \psi^c_t
\,,
\ee
and satisfies
\be
\label{Gstress}
\begin{array}{lcl}
{\displaystyle [G_r, J^a_n]} & = & {\displaystyle -n \psi^a_{r+n}} 
\vspace*{0.1cm} \\
{\displaystyle \{G_r, \psi^a_s\}} & = & {\displaystyle 
J^a_{r+s}\,.} 
\end{array}
\ee
The supersymmetric central charge is usually defined by
$$
\hat{c}={2\over 3} c =  { 3 k + 2 \over k} \,.
$$
The modes $L_n$ and $G_r$ satisfy the $N=1$ supersymmetric Virasoro
algebra
\be
\label{supervir}
\begin{array}{lcl}
{\displaystyle [L_m,L_n]} & = & {\displaystyle (m-n) L_{m+n}
+ {c \over 12} m (m^2 -1) \delta_{m,-n} } \vspace*{0.1cm} \\
{\displaystyle  [L_m,G_r]} & = & {\displaystyle 
\left({m \over 2} - r \right) G_{m+r}} \vspace*{0.1cm} \\
{\displaystyle  \{G_r,G_s\}} & = & {\displaystyle 
2 L_{r+s} + {c \over 3} \left(r^2 - {1\over 4} \right) \delta_{r,-s}}
\,. 
\end{array}
\ee
\medskip

As before we want to consider the case of a theory whose chiral algebra
is generated by two commuting subalgebras, where one subalgebra is the
above supersymmetric Kac-Moody algebra and the other defines a
(supersymmetric) unitary conformal field theory. The Virasoro
generators of the whole theory are then of the form 
$L_n=L_n^{su(1,1)} + L_n^0$, where $L_n^{su(1,1)}$ and $L_m^0$
commute, and the total central charge is 
\be
c = c^{su(1,1)} + c^0 = 15 \,.
\ee
The physical states are those states that satisfy
\be
L_n \phi = G_r \phi = 0 \qquad \hbox{for} \qquad n,r> 0 \,,
\ee
together with the mass-shell condition
\be
\label{msNS}
L_0 \phi = {1\over 2} \phi\,.
\ee
If the ground states transform in a representation of $su(1,1)$ whose
Casimir takes the value $-j(j+1)$ then the mass-shell condition
implies (as $h^0\geq 0$)
\be
- {j (j+1) \over 2k} +N \leq {1 \over 2} \,.
\ee
It is then again clear that for the continuous representations (where
$-j(j+1)=1/4 + \kappa^2$) only the ground states can satisfy the mass
shell condition, and the corresponding states have positive norm. The
only interesting cases are therefore the discrete representations
$D^{\pm}_j$. In the following we shall analyse in detail the case of 
$D^-_j$; the situation for $D^+_j$ is completely analogous.

In this section we want to show that the physical states in the Fock
space whose ground states transform in the $D^-_j$ representation of
$su(1,1)$ have positive norm provided that\footnote{This is slightly
stronger than the statement in \cite{Hwang2}.}
\be
0 > j > -k-1 \,.
\ee
The argument will be very similar to the argument in the
bosonic case. Let us denote by $\F$ the subspace of the Fock space
$\H$ that consist of states $\phi\in\H$ for which
\be
J^3_n \phi = L_n \phi = 0 \quad \hbox{for $n>0$} \quad
\psi^3_r \phi = G_r \phi = 0 \quad \hbox{for $r>0$,}
\ee
and denote by $\H^{(N)}$ the subspace of the Fock space that consists
of states whose grade is less or equal to $N$. In a first step we
prove the  
\bigskip

\noindent {\bf Lemma}. If $c=15$ and $0>j>-k-1$, then the states of
the form 
$$
|\{\varepsilon,\lambda,\delta,\mu\},f\rangle : = 
G_{-1/2}^{\varepsilon_1} \cdots G_{-a+1/2}^{\varepsilon_a}\,
L_{-1}^{\lambda_1} \cdots L_{-m}^{\lambda_m} \hspace*{4cm} 
$$
\be 
\label{basisNS}
\hspace*{3cm}
(\psi^3_{-1/2})^{\delta_1} \cdots (\psi^3_{-a+1/2})^{\delta_a} \,
(J^3_{-1})^{\mu_1} \cdots (J^3_{-m})^{\mu_m} |f\rangle \,,
\ee
where $f\in\F$ is at grade $L$, $\varepsilon_b,\delta_b\in\{0,1\}$,
and $\sum_b \varepsilon_b (b-1/2) + \sum_c \delta_c (c-1/2) 
+ \sum_r r \lambda_r + \sum_s s \mu_s +L \leq N$, form a basis for
$\H^{(N)}$. 
\bigskip

\noindent {\bf Proof}. Let us define 
\be
\label{vir3}
L^3_n = - {1 \over 2k} \sum_m : J^3_{n-m} J^3_m : \,,
\ee
and
\be
\label{g3}
G^3_r = - {1 \over k} \sum_s J^3_{r-s} \psi^3_s \,,
\ee
which satisfy the $N=1$ supersymmetric algebra $(\ref{supervir})$ with
$c=3/2$ ($\hat{c}=1$), and (\ref{stress}) and (\ref{Gstress}),
respectively, for $a=3$. We can then define  
\be
L^c_n = L_n - L^3_n \qquad G^c_r = G_r - G^3_r \,,
\ee
and, by construction, $L^c_n$ and $G^c_r$ commute (or anticommute)
with $J^3_n$ and $\psi^3_r$, and therefore with $L^3$ and $G^3$. This
implies that $L^c$ and $G^c$ define a $N=1$ supersymmetric algebra 
$(\ref{supervir})$ with $c=27/2$ ($\hat{c}=9$). 

Using (\ref{vir3}) and (\ref{g3}), we can rewrite the states in
(\ref{basisNS}) in terms of states where $L_n$ is replaced by $L_n^c$
and $G_r$ by $G_r^c$, and it is clear that this transformation defines
an isomorphism of vector spaces. In a first step we want to prove that
the states of the form (\ref{basisNS}) are linearly independent, and
to this end it is sufficient to do this for the modified states. As
$L^c$ and $G^c$ commute (or anticommute) with $J^3$ and $\psi^3$, the
Kac-determinant is then a product of the Kac-determinant corresponding
to the supersymmetric $U(1)$ theory (which is always non-degenerate),
and the Kac-determinant of a supersymmetric Virasoro highest
weight representation with $\hat{c}=9$ and highest weight
\be
h^c = h_f + {m^2 \over 2k} \,,
\ee
where $h_f$ and $m$ are the $L_0$-eigenvalue and $J^3_0$-eigenvalue of
the corresponding ground state $|f\rangle$. If $f$ is at grade $M$,
then
\be
\label{eqn}
h^c  = { - j^2 - j + m^2 + 2 M k \over 2k} + h^0 \,,
\ee
where $h^0\geq 0$ is the eigenvalue of $|f\rangle$ with respect to 
$L^0_0$. It is known that the degenerate representations at
$\hat{c}=9$ only arise for $h\leq 0$ \cite{cft}, and it therefore
remains to show that the first term is always positive.

For $M=0$, $m\leq j$, and (\ref{eqn}) is clearly positive, and for
$M=1/2$, $m\leq j+1$, and the numerator of the first term in
(\ref{eqn}) is bounded by $j+1+k>0$. For $M\geq 1$, we observe that
the possible values of $m$ are bounded by 
$m\leq j+M+1/2$, and it is therefore useful to consider the two cases 
(I) $j+M+1/2 < 0$, and (II) $j+M+1/2\geq 0$ separately. In case (II),
(\ref{eqn}) is minimal for $m=0$, and we can rewrite the numerator of
the first term on the right-hand side as 
\be
- j^2 - j + 2 M k  = - j (j+k+1) + k (2M+j) \,.
\ee
The first term is strictly positive for $0>j>-k-1$, and the second
term is non-negative (as for $M\geq 1$, $2M\geq M+1/2$). 

In case (I), (\ref{eqn}) is minimal for $m=j+M+1/2$, and then the
numerator of the first term on the right-hand side simplifies to
\be
-j^2 -j + (M+j + 1/2)^2 + 2 M k = 2 M (j+k+1) + (M-1/2)^2 \,.
\ee
This is also strictly positive, and we have thus shown that the states
of the form (\ref{basisNS}) are linearly independent. 
\medskip

We can then follow the same argument as in the Lemma of the previous
section to show that the states of the from (\ref{basisNS}) span the
whole Fock space. This completes the proof of the Lemma.
\bigskip

Let us call a state spurious if it is a linear combination of states
of the from (\ref{basisNS}) for which $\lambda\ne 0$ or
$\varepsilon\ne 0$. Because of the Lemma, every physical state $\phi$
can be written as a spurious state $\phi_s$ plus a linear combination
of states of the form (\ref{basisNS}) with $\lambda=0$ and
$\varepsilon=0$, {\it i.e.}
\be
\phi = \phi_s + \chi \,.
\ee
For $c=15$, following the argument of Goddard and Thorn \cite{GT}, 
$\phi_s$ and $\chi$ are separately physical states, and $\phi_s$ is
therefore null. Next we want to prove the 
\bigskip

\noindent {\bf Lemma.} Let $0>j>-k-1$. If $\chi$ is a physical state
of the form (\ref{basisNS}) with $\lambda=0$ and $\varepsilon=0$, 
then $\chi\in\F$. 
\medskip

\noindent {\bf Proof.} For fixed $|f\rangle$, let us denote by 
$\H_f$ the Fock space that is generated by the action of $J^3$ and
$\psi^3$ from $|f\rangle$, and by $\H_f^{svir}$ the Fock space that is
generated by the action of $L^3$ and $G^3$ from $|f\rangle$. Because
of (\ref{vir3}) and (\ref{g3}), it is clear that $\H^{svir}_f$ is a
subspace of $\H_f$. On the other hand $\H^{svir}_f$ is the Verma module
for the $N=1$ superconformal algebra with $c=3/2$ whose ground state has
conformal weight $- m^2 /2k$ where $m$ is the $J^3_0$ eigenvalue of
$|f\rangle$. It then follows from the Kac-determinant formula that
$\H^{svir}_f$ does not contain any null states unless $m=0$
\cite{cft}. Provided that $m\ne 0$, it is easy to see that
$\H^{svir}_f$ and $\H_f$ contain the same number of states at each
grade, and this implies that $\H_f=\H^{svir}_f$. Since $\H^{svir}_f$
does not contain any null states (with respect to the superconformal
algebra), it then follows that $\H_f$ does not contain any physical
states other than possibly $|f\rangle$ itself. It therefore remains to
check whether there are physical states with $m=0$, and if so whether
they are in $\F$. 
\smallskip

The physical states at fixed grade $N$ form a representation under the
zero mode $su(1,1)$ algebra since $J^{\pm}_0 \phi$ and $J^3_0\phi$ are
physical states if $\phi$ is. If the ground states form a
representation $D^-_j$ of the $su(1,1)$ zero mode algebra, then the
possible representations at grade $N$ are of the type $D_J^-$, where
$J$ is at most $j+N+1/2$. Because of the restriction on $j$, the
mass-shell condition implies 
\be
\label{massf}
N \leq {1 \over 2} + {j(j+1) \over 2k} < {1 \over 2} - {j\over 2}
< {1 \over 2} + {k+1 \over 2} \,.
\ee
For $0>j>-1$, the first inequality implies $N=0$, and then 
$m\leq j<0$. For $j=-1$, $N=0$ and $N=1/2$ are allowed; all of the
corresponding physical states satisfy $m<0$, except the state
(\ref{a1}) in appendix~A for which  $m=0$ (and $J=0$, $h^0=0$). This
state is however clearly in $\F$. For $-1>j>-2$ it follows that 
$N\leq 1$, since if $N\geq 3/2$, then $k>1$ by the last bound in 
(\ref{massf}). Hence $j(j+1)/2k<1$, but this contradicts the first
inequality in (\ref{massf}). Thus $m\leq j+1<0$.

Finally, for $j\leq -2$, then (\ref{massf}) implies directly that 
$N+j+1/2<1+j/2\leq 0$, and thus again $m\leq N+j+1/2<0$. This proves
the claim.
\bigskip

\noindent We are now ready to prove
\bigskip

\noindent {\bf Theorem}. For $c=15$ and $0>j>-k-1$, every physical
state $\phi$ differs by a spurious physical state from a state in
$\F$. Consequently, the norm of every physical state is non-negative. 
\bigskip

\noindent {\bf Proof.} This follows directly from the previous two
lemmas, and the fact that the coset $su(1,1)/u(1)$ 
is unitary if $0>j>-\tilde{k}$, as can be established by a slight
modification of the argument in \cite{DPL}. (The Kac-determinant of
the full Fock space is the product of the expression \cite[(4.8)]{DPL}
with $k$ replaced by $\tilde{k}$ and the fermionic
contributions. Apart from the fermionic part (which is manifestly
positive), the Kac-determinant of the coset model is then given by
\cite[(4.9)]{DPL}, where all $k$'s are replaced by $\tilde{k}$'s, except
for the $k$ in the factor $k^{-r_2(N)}$. This determinant is then
positive for $0>j>-\tilde{k}$ by the same arguments as in \cite{DPL}.)
\bigskip

Again, it is easy to see how to generalise the above argument to other
backgrounds (including the flat case). It is also clear by
continuity, as in the bosonic case, that the representations with 
$0> j \geq -k-1$ do not contain any physical states of negative norm. 
Furthermore, there always exist states of negative norm if this
condition is violated; we give examples in appendix~A. The analysis
for the representations $D^+_j$ is similar, and we find that all
physical states have positive norm provided that $0> j \geq -k-1$.

\section{The relation to the conformal field theory bound}

We now return to the conjectured relation between
type IIB string theory on ${\rm AdS}_3\times S^3\times M^4$ (where
$M^4$ is either K3 or $T^4$) and two-dimensional conformal field
theory  whose target is a symmetric product of a number of copies of
$M^4$ \cite{M}.  
This relation can be understood
by considering the string theory in the background of $Q_1$ D-strings
and $Q_5$ D5-branes. The theory on the world-volume of the D-strings
is a conformally-invariant sigma-model that has, 
in a certain limit, target space 
$Sym_{Q} M^4$, where $Q=Q_1 Q_5$ for $M^4 = T^4$ 
and $Q = Q_1Q_5 + 1$ for $M^4$ = K3 \cite{BSV}.

By S-duality, the background of the D1-D5 system is related to a
conventional IIB string theory on $SU(1,1)\times SU(2) \times M^4$,
where the level of the two WZW models is the same so that the total
central charge of the six-dimensional part of the theory is indeed
\be
c= c_{su(2)}(k) + c_{su(1,1)}(k) = 
{3 \over 2} \left( {3k -2 \over k} + {3k+2 \over k}\right) = 9\,.
\ee
According to \cite{MS,OV,CT}, the level of the $SU(1,1)$ and the
$SU(2)$ WZW model is $Q_5$, and one may therefore think that $k=Q_5/2$
(taking into account that $k$ is half-integral, and $Q_5$
integral). However, this assignment is somewhat delicate, as the $Q_1$
D-strings are mapped to $Q_1$ fundamental strings in the dual theory, 
and this interpretation is therefore only simple for $Q_1=1$, in which
case $k=Q/2$.  Nevertheless, for more general $Q_1>1$ one should
anticipate that the bound on the allowed values of the $U(1)$ charge
will be $Q_1$ times what it is for $Q_1=1$, and this means that the
effective level is again $k=Q/2$. This is the assumption we shall make
in extending our analysis from $Q_1=1$ to $Q_1>1$.

The superconformal field theory on $Sym_{Q} M^4$ has a $(4,4)$
superconformal algebra with $c=6 Q$ \cite{SV}. The level of the
$su(2)$ subalgebra is then $\ell = Q/2$ (in our conventions where the
level is half-integral) \cite{SS}, and the possible values of the $U(1)$
charge of primary $su(2)$ highest weight fields are therefore
$m=0,1/2,\ldots,\ell$. The primary fields that are chiral with respect to
a $N=2$ subalgebra (and that correspond to the BPS states of the dual
string theory) satisfy in addition $h=m$ \cite{LVW}.\footnote{Here we
have taken into account that in the usual conventions, the $U(1)$
generator of the $N=2$ subalgebra is twice the $T^3$ generator of the
$su(2)$ algebra of the $N=4$ algebra \cite{SS}.} There are therefore
only finitely many (namely $2 \ell + 1 = Q+1$) different chiral
primary fields, and this must thus be reflected in the dual string
theory; this is the content of the ``stringy exclusion principle'' of
Maldacena and Strominger \cite{MS}. 

In terms of the string theory on $SU(1,1)\times SU(2)\times M^4$, the
different values of $h(=m)$ are to be identified with the different
values of $-j$, the eigenvalue of $J^3_0$ of a $su(1,1)$ highest
weight in the $D^+_j$ representation \cite{MS}. The above bound
then transforms into the condition that $0\geq j \geq -\ell$. As is
explained in \cite{MS} (see \cite{BF}), a stability analysis on
${\rm AdS}_3$ suggests that $j\leq -1/2$. The above bound (together
with the stability bound) therefore gives 
$Q /2 \geq -j \geq 1/2$. For the case of K3,
$Q/2 = k + 1/2$, and we therefore obtain precisely the range of
allowed representations $k+1 > -j > 0$ which we have shown to be
ghost-free.\footnote{The only other ghost-free representation occurs
for $-j = k+1$ but differs from the others in that it contains null
vectors; it presumably does not occur in a modular-invariant partition
function \cite{Hwang3}.}
In the case of $T^4$, however, we obtain 
$k > -j > 0$, which is a more restrictive condition, corresponding to
a proper subset of the ghost-free representations.
A priori we have no grounds for expecting the two restrictions to
coincide except in the limit of large $Q$.
It is encouraging that this is indeed what occurs, and particularly
interesting that the bounds coincide exactly for the case of K3.

\section{Conclusions}
\setcounter{equation}{0}

In this paper we have analysed the no-ghost theorem for string theory
on $SU(1,1)$. We have filled the gap in the proof of \cite{Hwang4} in the
bosonic case, and extended the argument to the fermionic case. We have
also shown that the restriction on the spin (in terms of the level)
that is necessary to obtain a ghost-free spectrum corresponds to the
stringy exclusion principle of Maldacena and Strominger \cite{MS}.
Among other things, we regard this is as evidence that the $SU(1,1)$
model with the restriction on the set of allowed representations
defines a consistent string theory.

There are many interesting questions which need to be addressed. In
order to get a consistent string theory the amplitudes must be
crossing symmetric, and it is not clear whether this can be achieved
with the restricted set of representations. This is a rather difficult
problem as the fusion rules of the $SU(1,1)$ WZW model are not well
understood  (see however recent progress on an understanding of
the fusion rules of the $SU(2)$ WZW model at fractional level which is
technically similar \cite{GW}). Furthermore, in order to get a modular
invariant theory, additional representations (that correspond to
winding states along the compact direction in $SU(1,1)$) presumably
have to be considered \cite{Hwang3,Hwang5}, for which the $L_0$
spectrum is not bounded from below. Finally, the set of ghost-free
representations contains a continuum, the so-called continuous
representations of the  global $SU(1,1)$, and thus problems similar to
those faced in Liouville theory \cite{ZZ} arise. Nevertheless, it is
quite suggestive that the representations that are allowed by
the no-ghost-theorem are those representations whose Verma 
module does not contain any null-vectors \cite{Hwang5},\footnote{This
is the case for the continuous representations, and for the discrete
representations $D^-_j$ if we impose the strict inequality $0>j>-k$ in
the bosonic case (and $0> j > -k-1$ in the fermionic case) and
similarly for $D^+_j$; strictly speaking the no-ghost-theorem allows
also $j\geq -k$ and $j\geq -k-1$, respectively.} and this may
ultimately be sufficient to prove that the restricted representations
define a consistent interacting theory. One may also hope that the
structure of the dual superconformal field theory could shed light on
some of these questions.

\section*{Acknowledgements}

We thank Michael Green for important suggestions and encouragement. We
also thank Ingo Gaida, Peter Goddard, Adrian Kent, Hugh Osborn and
Mark Walton for helpful comments.  
\smallskip

J.M.E. is grateful to PPARC for an Advanced Research Fellowship, and 
M.R.G. is supported by a Research Fellowship of Jesus College,
Cambridge.

\section*{Appendix}

\appendix

\section{Some illustrative examples for the supersymmetric case}
\setcounter{equation}{0}

Let us determine the norms of the physical states at the various
grades. At every grade, we shall look for physical highest weight
states that generate the representation $D^-_J$.

\subsection*{Grade 1/2}

${\bf J}=j+1$: There is one physical state 
\be
\label{a1}
P_{\half,1}:=\psi^+_{-1/2} |j,j\rangle \qquad || P_{1,1} ||^2 = 2 \, k \,.
\ee
This state has positive norm.
\bigskip

\noindent ${\bf J}=j$: A physical state only exists for $j=-1$, in
which case it is given as 
\be
P_{\half,0}:= \left(\psi^3_{-1/2} + {1 \over 2} 
\psi^+_{-1/2} J^-_{0} \right) |-1,-1\rangle \,.
\ee
The norm of this state is $0$.
\bigskip

\noindent ${\bf J}=j-1$: There is one physical state of the form
\be
P_{\half,-1}:= \left( \psi^-_{-1/2} - {1 \over j} \psi^3_{-1/2} J^-_0 
+ {1 \over 2 \, j \, (2\, j -1)} \psi^+_{-1/2} J^-_0 J^-_0 \right)
|j,j\rangle \,,
\ee
and its norm square is 
\be
|| P_{\half,-1} ||^2 = 2 \, k { (2\, j+1) \over (2\, j -1)} \,.
\ee
This is positive as the mass-shell condition implies $j\leq -1$.

\subsection*{Grade 1}

\noindent ${\bf J}=j+1$: There is one physical state of the form 
\be
P_{1,1}:= \left( J^+_{-1} 
+ {1 \over j+1} \psi^+_{-1/2} \psi^3_{-1/2} \right) |j,j \rangle \,,
\ee
whose norm square is
\be
|| P_{1,1} ||^2 = 2 {(k+j+1) (j(j+1)-k) \over (j+1)^2} \,.
\ee
The second bracket in the numerator is non-negative because of the
mass-shell condition (\ref{msNS}) at grade $N=1$, and the expression
is therefore non-negative if and only if $j\geq -k-1$ holds. 
\bigskip

\noindent ${\bf J}=j$: There is one physical state of the form
\be
P_{1,0}:= \left( \psi^+_{-1/2} \psi^-_{-1/2} 
-  {1 \over j} \psi^+_{-1/2} \psi^3_{-1/2} J^-_0 \right) |j,j\rangle
\,,
\ee
and its norm square is 
\be
|| P_{1,0} ||^2 = 4 k^2 {j+1 \over j} \,.
\ee
This is positive as the mass-shell condition implies $j<-1$.
\bigskip

\noindent ${\bf J}=j-1$: There is one physical state whose norm square
is 
\be
||P_{1,-1}||^2 = -2 {(2\, j+1) (k-j) (k - j(j+1)) \over j^2 (2\,j-1)}
\,. 
\ee
Because of the mass shell condition (\ref{msNS}) with $N=1$, the last
bracket in the numerator is non-positive and $j\leq -3/2$. Thus the
norm is non-negative.

\subsection*{Grade 3/2}

\noindent ${\bf J}=j+2$: There is one physical state of the form 
\be
P_{\thalf,2} = \psi^+_{-1/2} J^+_{-1} |j,j\rangle \,,
\ee
whose norm square is 
\be
|| P_{\thalf,2} ||^2 = 4 k (j+k+1) \,.
\ee
This is non-negative if $j\geq -k-1$.
\bigskip

\noindent ${\bf J}=j+1$: There is one physical state whose norm square
is 
\be
|| P_{\thalf,1} ||^2 = - 2 \, j \, k \, (j+2) \, (2k - j\, (j+1)) \,.
\ee
This is non-negative since $j\leq -2$ (for $j=-3/2$ only $N=1$ is
possible), and $2k - j (j+1)\leq 0$ because of (\ref{msNS}) with
$N=3/2$.  
\bigskip

\noindent ${\bf J}=j$: There is a two-dimensional space of physical
states. The determinant of the $2\times 2$ inner product matrix is 
$$
\hbox{Det} = - 64 \, k^2 \, j \,  (2\, j+3) \, (2\, j-1) \, (j+1) 
\, (j+k+1) \hspace*{4cm} 
$$
$$
\hspace*{2cm} \times \Bigl(2\,k - j\,(j+1)\Bigr)\, (k-j) \,
{ k\, (3\, k+2) + j \, (j+2)\, (j+1)\, (j-1) \over 
(3\,k+j\,(j+1))^2} \,,
$$
which is manifestly positive. As the two eigenvalues are positive for
large $j$ and $k$, the only negative norm states can occur if the
determinant vanishes, which can happen for $k=j(j+1)/2$ and
$k=-1-j$. In the former case, the trace of the inner product matrix is
then 
\be
\hbox{Trace} (k=j(j+1)/2) = {2 \over 25} (2j+3) (j+2) (j+1) (j-1)
(2j-1) j (7j(j+1) - 4) \,,
\ee
which is non-negative for $j\leq -2$, and in the second case the trace
is  
\be
\hbox{Trace} (k=-j-1) =  4 (16 j(j+1) + 5) 
{j (2j-1) (j+2) (j+1) (j^2 +1) \over (j-3)^2} \,,
\ee
which is also non-negative for $j\leq -2$. This demonstrates that there
are no negative norm states in this case.

\end{document}